\documentstyle[preprint,aps,a4,12pt]{revtex}
\begin{document}
\draft
\hfill\vbox{\baselineskip14pt
            \hbox{\bf KEK-TH-560}
            \hbox{KEK Preprint xx-xx}
            \hbox{January 1998}}
\baselineskip20pt
\vskip 0.2cm 
\begin{center}
{\Large\bf The Relevant Operators for the Hubbard Hamiltonian
with a magnetic field term}
\end{center} 
\vskip 0.2cm 
\begin{center}
\large S.~Alam\footnote{Permanent address: Department of Physics, University
of Peshawar, Peshawar, NWFP, Pakistan.}, M.~Nasir~Khan
\footnote{Institute of Applied Physics, Tsukuba University, 
Tsukuba, Ibaraki 305, Japan.}
Jauhar Ali\footnote{Institute of Information Sciences and Electronic,
University of Tsukuba, Tsukuba, Ibaraki 305, Japan.}
\end{center}
\begin{center}
{\it Theory Group, KEK, Tsukuba, Ibaraki 305, Japan }
\end{center}
\vskip 0.2cm 
\begin{center} 
\large Abstract
\end{center}
\begin{center}
\begin{minipage}{14cm}
\baselineskip=18pt
\noindent
The Hubbard Hamiltonian and its variants/generalizations
continue to dominate the theoretical modelling of important
problems such as high temperature superconductivity. 
In this note we identify the set of relevant operators
for the Hubbard Hamiltonian with a magnetic field term.
\end{minipage}
\end{center}
\vfill
\baselineskip=20pt
\normalsize
\newpage
\setcounter{page}{2}
\section{Introduction}
	The Hubbard Hamiltonian [HH] and its extensions dominate
the study of strongly correlated electrons systems and the 
insulator metal transition \cite{fra91}. One of the attractive
feature of the Hubbard Model is its simplicity. It is well known
that in the HH the band electrons interact via a two-body
repulsive Coulomb interaction; there are no phonons in this model
and neither in general are attractive interactions incorporated.
With these points in mind it is not surprising that the HH
was mainly used to study magnetism. In contrast superconductivity
was understood mainly in light of the BCS theory, namely as
an instability of the vacuum [ground-state] arising from
effectively attractive interactions between electron and 
phonons. However Anderson \cite{and87} suggested that
the superconductivity in high T$_{c}$ material could arise
from purely repulsive interaction. The rationale of
this suggestion is grounded in the observation that
superconductivity in such materials arises from the
doping of an otherwise insulating state. Thus following this
suggestion the electronic properties in such a 
high T$_{c}$ superconductor material close to a 
insulator-metal transition must be considered.
In particular the one-dimensional HH is considered
to be the most simple model which can account
for the main properties of strongly correlated
electron systems including the metal-insulator 
transition. Long range anti-ferromagnetic order
at half-filling has been reported in the numerical
studies of this model \cite{hir89,bal90}.
Away from half-filling this model has been studied
in \cite{mor90,mor91}. 

	The Maximal Entropy Principle [MEP] is a useful tool
to get the dynamical and thermodynamical descriptions.
The main advantage of this formalism is to provide
a definite prescription to determine the complete
set of operators [i.e. relevant operators] related
to the problem under considerations. An attractive
feature of the relevant operators is that they are
group theory based and hence once a Hamiltonian
for a system can be written down, the task of
identifying the relevant operators can proceed
in principle. Relevant operators for a given 
physical system along with the Hamiltonian
in essence describe the essential bare bones
of the physical system. 
In a series of papers \cite{ali91,gru93,gru95}, the generalized
time-dependent Jaynes-Cummings Hamiltonian in the context of Maximum
Entropy Principle [MEP] and group theory based methods \cite{pran90} 
was studied. In particular, in \cite{ali91} the MEP formalism was 
used to solve time-dependent N-level systems. A set of generalized 
Bloch equations, in terms of relevant 
operators was obtained and as an example the $N=2$ case was solved.
It was thus demonstrated in \cite{ali91} that the dynamics and 
thermodynamics of a two-level system coupled to a classical field 
can be fully described in the framework of MEP and group theory
based methods. 
Further in \cite{gru93} a time-dependent generalization of the JCM
was studied and by showing that the initial conditions of the operators
are determined by the MEP density matrix the authors were able to
demonstrate that inclusion of temperature turns the problem into a
thermodynamical one. An exact solution was also presented in the
time independent case. Finally in \cite{gru95} more detailed analysis
of the three set of relevant operators was given. These set of
operators are related to each other by isomorphisms which allowed
the authors to consider the case of mixed initial conditions.
	The mean values of the field's population, correlation
functions and $n$th-order coherence functions are of interest
and useful in several applications.
The MEP formalism allows us to describe a Hamiltonian system
in terms of those, and only those, quantum operators 
{\it relevant} to the problem at hand. Thus, this formalism 
is suitable to study the Hamiltonian given 
in \cite{gru93,gru95}. In \cite{gru93,gru95}
the {\em population of each level and not their difference} 
is considered therefore the resulting Hamiltonian is 
called a {\em generalized} time-dependent JCH.
Recently the relevant operators for the generalized
time-dependent m-photon Jaynes-Cummings
Hamiltonian were determined in \cite{ala97}.

	The HH for different band-fillings is studied in the 
context of MEP by Aliga and Proto \cite{ali92}. The
HH with a magnetic field was considered by Alam and
Proto \cite{ala93} using MEP techniques. In the present 
note we incorporate a magnetic field term in HH and
identify the relevant operators. The set of relevant
operators and their evolution equations without
the magnetic field term considered by Aliga and
Proto has also been independently checked by us.
Moreover by neglecting the magnetic field term
we easily recover the case considered by Aliga
and Proto which provides a check on our calculations.
It is interesting to note that Essler et al. 
\cite{ese92,ese93} have suggested an extended 
Hubbard model which contains the t-J model as
a special case. In fact this model is a mixture
of the Hubbard and the t-J model. The model of 
Essler et al. \cite{ese92} contains
a magnetic field term. On a one-dimensional
lattice Essler et al.~\cite{ese92} present
an exact solution to their model via Bethe
ansatz. It is further claimed that by using
$\eta$-pairing mechanism one can construct
eigenstates of the Hamiltonian with off-diagonal
long-range order and that in the attractive case
the exact ground state is superconducting
in any numbers of dimensions. The model
of Essler et al.~\cite{ese92,ese93} is 
motivated by high-T$_{c}$ superconductivity 
and is expected to describe a system of strongly
correlated electrons. The model of Essler 
et al.\cite{ese92} possesses a huge symmetry
group [for example it has eight supersymmetries]
and it would be interesting to obtain the set
of relevant operators corresponding to it.  
 
The main purpose of this paper is to answer the question:
Can we identify a set of relevant operators for the 
Hubbard Hamiltonian including a magnetic field term? 
The aims of this short note is to give
such a set and the evolution equations for it.
The layout of this paper is as follows. Section two contains
discussion and definitions relevant to Hubbard Hamiltonian
in the context of mean field method. In section three we
recall some well-known results of the group theory based MEP
formalism. In section four we give the relevant operators and
the evolution equations for their expectation values in the
context of the Hubbard Hamiltonian without and with an
a magnetic field term present. Conclusions are given
in the last section.  

\section{Hubbard Hamiltonian and the Mean Field Method}
The Hubbard Hamiltonian can be written as
\begin{equation}
\hat{H}=-\tau^{'}\sum_{<i,j>,~\sigma}\hat{c}^{\dagger}_{i\sigma}
\hat{c}_{j\sigma}+U\sum_{i}\hat{n}_{i\uparrow}
\hat{n}_{i\downarrow}.
\label{H1}
\end{equation}
$\tau^{'}$ is the hopping parameter between the nearest
neighbours, $\hat{c}^{\dagger}_{i\sigma}$
creates an electron with spin $\sigma$ at site $i$, 
$\hat{c}_{i\sigma}$ destroys an electron with spin $\sigma$ at 
site $i$, $U$ is the on-site Coulomb interaction and
$\hat{n}_{i\sigma}=\hat{c}^{\dagger}_{i\sigma}\hat{c}_{i\sigma}$ 
is the number operator for spin $\sigma$ at site $i$.
 
	A modified Hubbard Hamiltonian,
\begin{equation}
\hat{H}_{1}=-\tau^{'}\sum_{<i,j>,~\sigma}\hat{c}^{\dagger}_{i\sigma}
\hat{c}_{j\sigma}+U\sum_{i}[\hat{n}_{i\uparrow}-\frac{1}{2}]
[\hat{n}_{i\downarrow}-\frac{1}{2}]],
\label{H2}
\end{equation}
is also used by some authors \cite{mor90,mor91}.
One may rewrite $\hat{H}_{1}$ in terms of $\hat{H}$
\begin{eqnarray}
\hat{H}_{1}&=&-\tau^{'}\sum_{<i,j>,~\sigma}\hat{c}^{\dagger}_{i\sigma}
\hat{c}_{j\sigma}+U\sum_{i}[\hat{n}_{i\uparrow}\hat{n}_{i\downarrow}
-\frac{1}{2}(\hat{n}_{i\uparrow}+\hat{n}_{i\downarrow})+\frac{1}{4}]\nonumber\\
    &=& H-\frac{U}{2}\sum_{i}[\hat{n}_{i\uparrow}+\hat{n}_{i\downarrow}]
+\frac{1}{4}N U \hat{I}.
\label{H3}
\end{eqnarray}
$N$ in Eq.~\ref{H3} is the number of sites.
It is important to note that the Hamiltonians
given in Eq.~\ref{H1} and Eq.~\ref{H3} cannot
be considered as equivalent even when they lead
to the same set of the relevant operators, since
the $g$ matrix [see Eq.~\ref{F5} below] associated
with the Hamiltonian in Eq.~\ref{H1} is different
from that which corresponds to the Hamiltonian in 
Eq.~\ref{H3}. 

In order to get solvable model, in this note, we resort
to the mean-field method. Our main approximation is
to replace the product of operators in the hopping
term by averages according the rule
\begin{equation}
\hat{A}\hat{B}=<\hat{A}>\hat{B}+\hat{A}<\hat{B}>-<\hat{A}><\hat{B}>.
\label{H4}
\end{equation}
It is important to note that in contrast to \cite{don90} 
the mean-field approximation in our case,like \cite{all87},  
has been applied to the hopping term. In
\cite{don90} the mean-field approximation is applied
to the Coulomb term. 
In our approximation the Hamiltonian can be written
in site-diagonal form, the sites being coupled only 
by the mean-field parameter $\Delta_{\sigma}$, for
the definition of $\Delta_{\sigma}$, see Eq.~\ref{H7}
below.

In order to apply the above rule, viz Eq.~\ref{H4} to 
the hopping term we rewrite the latter as

\begin{eqnarray}
\sum_{<i,j>,~\sigma}\hat{c}^{\dagger}_{i\sigma}\hat{c}_{j\sigma}
&=&\hat{c}^{\dagger}_{i\sigma}\hat{c}_{j\sigma}+
\hat{c}^{\dagger}_{j\sigma}\hat{c}_{i\sigma},\nonumber\\
&=& [\hat{c}^{\dagger}_{i\sigma}+ \hat{c}^{\dagger}_{j\sigma}]
 [\hat{c}_{i\sigma}+ \hat{c}_{j\sigma}]
-\hat{n}_{i\sigma}-\hat{n}_{j\sigma},
\label{H5}
\end{eqnarray}

Applying the definition of the averaging procedure,
viz Eq.~\ref{H4} to the hopping term written as 
in Eq.~\ref{H5} we obtain

\begin{eqnarray}
\sum_{<i,j>,~\sigma}\hat{c}^{\dagger}_{i\sigma}\hat{c}_{j\sigma}
&=&\sum_{\sigma}\hat{c}^{\dagger}_{i\sigma}\hat{c}_{j\sigma}+
\hat{c}^{\dagger}_{j\sigma}\hat{c}_{i\sigma},\nonumber\\
&\approx& \sum_{\sigma}\Delta_{\sigma}^{*}
<\hat{c}^{\dagger}_{i\sigma}+ \hat{c}^{\dagger}_{j\sigma}>
+\Delta_{\sigma}<\hat{c}_{i\sigma}+ \hat{c}_{j\sigma}>\nonumber\\
&&-|<\Delta_{\sigma}>|^2 I-\hat{n}_{i\sigma}-\hat{n}_{j\sigma},
\label{H6}
\end{eqnarray}

where we have used the definitions 

\begin{eqnarray}
\Delta_{\sigma}&=& <\hat{c}^{\dagger}_{i\sigma}
+\hat{c}^{\dagger}_{j\sigma}>,\nonumber\\
\Delta_{\sigma}^{*}&=& <\hat{c}_{i\sigma}+\hat{c}_{j\sigma}>.
\label{H7}
\end{eqnarray}

Using the reduction given in Eq.~\ref{H6} we may
write the Hubbard Hamiltonian in Eq.~\ref{H1} as

\begin{eqnarray}
\hat{H}&=&\tau^{'}\sum_{\sigma}\hat{n}_{i\sigma}+\hat{n}_{j\sigma}
-\tau^{'}\sum_{\sigma}\Delta_{\sigma}[\hat{c}_{i\sigma}+ \hat{c}_{j\sigma}]
-\tau^{'}\sum_{\sigma}\Delta_{\sigma}^{*}
[\hat{c}^{\dagger}_{i\sigma}+ \hat{c}^{\dagger}_{j\sigma}]\nonumber\\
&&+\tau^{'}\sum_{\sigma}|<\Delta_{\sigma}>|^2 I 
+U\sum_{i}\hat{n}_{i\uparrow}\hat{n}_{i\downarrow}.
\label{H8}
\end{eqnarray}

Next we want to write the HH in terms of one site
only to this end we observe that in one dimension
[1-d] each site has two nearest neighbours, in
2-d each site as 4 nearest neighbours and so on.
Denoting the number of nearest neighbours by $m$ we define
\begin{eqnarray}
\tau &\stackrel{\rm{def}}{=} & m\tau^{'},\nonumber\\
|\Delta|^2 &\stackrel{\rm{def}}{=}& \frac{|\Delta_{\uparrow}|^2
+|\Delta_{\downarrow}|^2}{2} 
\label{H9}
\end{eqnarray} 

Using these definitions we may write the one-site 
equivalent of Eq.~\ref{H8}
\begin{eqnarray}
\hat{H}_{i}&=&\tau \sum_{\sigma}\hat{n}_{i\sigma}
-\tau\sum_{\sigma}\Delta_{\sigma}\hat{c}_{i\sigma}
-\tau\sum_{\sigma}\Delta_{\sigma}^{*}\hat{c}^{\dagger}_{i\sigma}\nonumber\\
&&+\tau |\Delta|^2 \hat{I}+U \hat{n}_{i\uparrow}\hat{n}_{i\downarrow}.
\label{H10}
\end{eqnarray}

Eq.~\ref{H10} allows us to rewrite the Hamiltonian
in Eq.~\ref{H3} in the form
\begin{eqnarray}
\hat{H}_{1i}&=&(\tau -\frac{U}{2})\sum_{\sigma}\hat{n}_{i\sigma}
-\tau\sum_{\sigma}\Delta_{\sigma}\hat{c}_{i\sigma}
-\tau\sum_{\sigma}\Delta_{\sigma}^{*}\hat{c}^{\dagger}_{i\sigma}\nonumber\\
&&+(\tau |\Delta|^2+\frac{U}{4})\hat{I} 
+U \hat{n}_{i\uparrow}\hat{n}_{i\downarrow}.
\label{H11}
\end{eqnarray}

	It is convenient to introduce a compact notation
\begin{eqnarray}
\hat{H}_{i}&=&\alpha \hat{n}_{i}-\tau \hat{x}_{i}
+\gamma \hat {I} +U \hat{r}_{i},
\label{H12}
\end{eqnarray}
where we have defined
\begin{eqnarray}
\hat{n}_{i} &\stackrel{\rm{def}}{=} &
 \hat{n}_{i\uparrow}+\hat{n}_{i\downarrow},\nonumber\\
\hat{x}_{i} &\stackrel{\rm{def}}{=} & 
\Delta_{\uparrow}^{*}\hat{c}^{\dagger}_{i\uparrow}
+\Delta_{\downarrow}^{*}\hat{c}^{\dagger}_{i\downarrow}
+\Delta_{\uparrow}\hat{c}_{i\uparrow}
+\Delta_{\downarrow}\hat{c}_{i\downarrow}
,\nonumber\\
\hat{r}_{i} &\stackrel{\rm{def}}{=} & 
\hat{n}_{i\uparrow}\hat{n}_{i\downarrow}.
\label{H13}
\end{eqnarray} 
$\hat{n}_{i}$ is the number of electrons at site i,
$\hat{x}_{i}$ is the mean field hopping interaction
between neighbouring sites and $\hat{r}_{i}$ measures
the double occupancy probability or simply the number
of pairs at the site i. 

	The Hamiltonian in Eq.~\ref{H10} is a
special case of the Hamiltonian form in 
Eq.~\ref{H12} with the identifications
$\alpha=\tau$ and $\gamma=\tau |\Delta|^2$.
If we set $\alpha=\tau-\frac{U}{2}$ and
$\gamma=\tau |\Delta|^2+\frac{U}{4}$ in
Eq.~\ref{H12} we recover the Hamiltonian
form given in Eq.~\ref{H11}.

	The magnetic field can be readily
accommodated by adding the term 
$h(\hat{n}_{i\uparrow}-\hat{n}_{i\downarrow})$
to the Hamiltonian form in Eq.~\ref{H12}, viz,
\begin{eqnarray}
\hat{H}_{i}&=&\alpha \hat{n}_{i}-\tau \hat{x}_{i}
+\gamma \hat {I} +U \hat{r}_{i}+h \hat{N}_{i},
\label{H14}
\end{eqnarray}
where we have defined
\begin{eqnarray}
\hat{N}_{i} &\stackrel{\rm{def}}{=} &
 \hat{n}_{i\uparrow}-\hat{n}_{i\downarrow}.
\label{H15}
\end{eqnarray}  
$\hat{n}_{i}$ and $\hat{N}_{i}$ respectively represent 
the symmetric and antisymmetric sums of the number
operators for both types of spin. We define 
$|\tilde{\Delta}|^2$ in analogy with $|\Delta|^2$
as
\begin{eqnarray}
|\tilde{\Delta}|^2 &\stackrel{\rm{def}}{=}& 
\frac{|\Delta_{\uparrow}|^2-|\Delta_{\downarrow}|^2}{2},\nonumber\\
 |\Delta_{\uparrow}|^2 &=& 
|\Delta|^2+\tilde{\Delta}|^2,\nonumber\\
|\Delta_{\downarrow}|^2 &=& |\Delta|^2-\tilde{\Delta}|^2,
\label{H16}
\end{eqnarray} 
where we have written $|\Delta_{\uparrow}|^2$ and
$|\Delta_{\downarrow}|^2$ in terms of  $|\Delta|^2$
and $|\tilde{\Delta}|^2$. 

\section{Outline of the MEP Formalism}
	It is instructive to summarize the principal 
concepts of the MEP \cite{ali91,gru93,gru95,alh77,pro89}.
A summary of MEP formalism has been given in 
\cite{ala97}. Here we again outline it for the benefit 
of the readers not familiar with \cite{ala97}.
 
Given the expectation values $<\! \hat O_{j} \!>$ of the operators
$\hat O_{j}$, the statistical operator $\hat \rho(t)$ is defined by
\begin{equation}
\hat \rho(t) = \exp\left(-\lambda_{0}\hat I-\sum_{j=1}^{L}\lambda
_{j}\hat O_{j}\right),\label{F1}
\end{equation}\\[0.3ex]
where $L$ is a natural number or infinity, and the $L+1$ Lagrange
multipliers $\lambda_{j}$, are 
determined to fulfill the set of constraints
\begin {equation}
<\! \hat O_{j} \!> = {\rm Tr} \;[\: \hat \rho(t) \;  \hat O_{j} \:] \; ,
\hspace{1.0cm} \mbox{j = 0, 1, \ldots, L} \; ,\label{F2} 
\end{equation}\\[0.3ex]
($\hat O_{0}=\hat I$ is the identity operator) and the normalization
in order to maximize the entropy, defined (in units of the Boltzmann
constant) by
\begin {equation}
S(\hat \rho) =-{\rm Tr}\;[\: \hat \rho  \;  \ln \hat \rho \:] \; .\label{F3}
\end{equation}\\[0.3ex]
Eq.~\ref{F1} is a generalization of the more familiar
density operator. For e.g. in open system, where we
have Grand Canonical Ensemble there are two Lagrange 
multipliers, $\beta=\frac{1}{k_B T}$ and $\mu$  are present, 
and we write the density operator as \cite{rei87} 
\begin{equation}
\hat \rho(t)=
\exp\left(\beta\Omega(T,V,\mu)-\beta \hat{H}+\beta\mu \hat{N}\right),
\label{F1x}
\end{equation}\\[0.3ex]
As is well-known the dynamics are governed by the time evolution 
of the  statistical operator.
The time evolution of the statistical operator is given by
\begin {equation}
i\hbar \frac{d \hat \rho}{dt}=[\: \hat H(t), \hat \rho(t) \:] 
\; .\label{F4}
\end{equation}\\[0.3ex]

The essence of the MEP formalism in conjunction with the
group theory method is to find the relevant operators 
entering Eq.~\ref{F1}) so as to guarantee not only 
that $S$ is maximum, but also is a constant of
motion. Introducing the natural logarithm of Eq.~\ref{F1} into 
Eq.~\ref{F4}) it can be easily verified that the 
{\bf relevant operators}
are  those that close a semi-Lie algebra under 
commutation with the Hamiltonian  $\hat H$, i.e.
\begin {equation}
[\: \hat H(t),\hat O_{j} \:]= i\hbar\sum_{i=0}^{L}g_{ij}(t)\hat O_{i} \; .
\label{F5}
\end{equation}\\[0.3ex]
Thus the relevant operators may be defined as those satisfying
the above equation.
Equation~\ref{F5}) defines an $L \times L$ matrix $G$ and 
constitutes the central requirement to be fulfilled by the 
operators entering in the density
matrix. The Liouville Eq.~\ref{F4} can
be replaced by a set of coupled equations for the 
mean values of the relevant operators or the Lagrange 
multipliers as follows \cite{ali89}:
 \begin {equation}
\frac{d<\! \hat O_{j} \!>_{t}}{dt} 
= -\sum_{i=0}^{L} g_{ij} <\! \hat O_{i} \!> \; ,
\hspace{1.0cm} \mbox{j = 0, 1, \ldots, L} \; ,\label{F6} 
\end{equation}
\begin {equation}
\frac{d\lambda_{j}}{dt} = \sum_{i=0}^{L} \lambda_{i} g_{ji},
\hspace{1.0cm} \mbox{j = 0, 1,\ldots, L}.\label{F7}
\end{equation}\\[0.3ex]
In the MEP formalism, the mean value of the operators 
and the Lagrange multipliers belongs to dual spaces 
which are related by \cite{pro89}
\begin{equation}
<\! \hat O_{j} \!>=-\frac{ \partial \lambda_{0} }{ \partial \lambda_{j}}.
\label{F8}
\end{equation}\\[0.3ex]
\section{The Relevant Operators and Evolution Equations
for the Hubbard Hamiltonian with a magnetic field term}
	For notational convenience we now drop the
subscript i, in all formulae from now on. The set
of relevant operators for the HH with a magnetic field 
term is more than twice
the number of relevant operators without it. It is
thus informative and useful to give the set of
the relevant operators for the HH without the magnetic
field. To this end we first consider the Hamiltonian form
given in Eq.~\ref{H12}. A little work shows that
number operator $n$ does not commute with the
Hamiltonian, after some calculation we obtain
\begin{equation}
\left[\hat{H},\hat{n}\right] = -i \tau \hat{p} ,
\label{R1}  
\end{equation}\\[0.3ex]
where $\hat{p}$ is given by
\begin{eqnarray}
\hat{p} &\stackrel{\rm{def}}{=} & 
i(\Delta_{\uparrow}^{*}\hat{c}^{\dagger}_{\uparrow}
+\Delta_{\downarrow}^{*}\hat{c}^{\dagger}_{\downarrow}
-\Delta_{\uparrow}\hat{c}_{\uparrow}
-\Delta_{\downarrow}\hat{c}_{\downarrow}).
\label{R2}
\end{eqnarray} 
$\hat{p}$ is the mean field electron's current.
 Thus so far we have introduced three operators besides
the Hamiltonian, namely $\hat{n}$, $\hat{x}$ and $\hat{p}$ 
belonging to the relevant operator set.
To determine the whole set we must proceed by
finding the commutation relations of all the
operators with the Hamiltonian until we
get the complete set. The commutation relation
of $\hat{x}$ with the Hamiltonian yields
\begin{equation}
\left[\hat{H},\hat{x}\right] = -i \alpha \hat{p}-iU\hat{l}_{-} ,
\label{R3}  
\end{equation}\\[0.3ex]
where $\hat{l}_{-}$ is the mean field pair's current
and can be written as
\begin{eqnarray}
\hat{l}_{-} &\stackrel{\rm{def}}{=} & 
i([\Delta_{\uparrow}^{*}\hat{c}^{\dagger}_{\uparrow}
-\Delta_{\uparrow}\hat{c}_{\uparrow}]\hat{n}_{\downarrow}
+\hat{n}_{\uparrow}[\Delta_{\downarrow}^{*}
\hat{c}^{\dagger}_{\downarrow}
-\Delta_{\downarrow}\hat{c}_{\downarrow}]).
\label{R4}
\end{eqnarray} 
We note that since the c's are fermion operators
they anticommute, hence as a consequence of this
$\hat{c}_{\downarrow}$ commutes with
$\hat{n}_{\uparrow}$, viz, explicitly,
$\hat{c}_{\downarrow}\hat{n}_{\uparrow}=
\hat{c}_{\downarrow}\hat{c}^{\dagger}_{\uparrow}
\hat{c}_{\uparrow}=-\hat{c}^{\dagger}_{\uparrow}
\hat{c}_{\downarrow}\hat{c}_{\uparrow}=
\hat{c}^{\dagger}_{\uparrow}\hat{c}_{\uparrow}
\hat{c}_{\downarrow}=\hat{n}_{\uparrow}\hat{c}_{\downarrow}$.

The commutation relation of $\hat{p}$ with the
Hamiltonian introduces yet another two operators
$\hat{l}_{+}$ and $\hat{\omega}_{1}$. $\hat{l}_{+}$
represents the mean field pair's interaction. 
\begin{equation}
\left[\hat{H},\hat{p}\right] = i \alpha \hat{x}+iU\hat{l}_{+} 
-i 4 \tau \hat{\omega}_{1},
\label{R5}  
\end{equation}\\[0.3ex]
where $\hat{l}_{+}$ and $\hat{\omega}_{1}$ are defined as
\begin{eqnarray}
\hat{l}_{+} &\stackrel{\rm{def}}{=} & 
([\Delta_{\uparrow}^{*}\hat{c}^{\dagger}_{\uparrow}
+\Delta_{\uparrow}\hat{c}_{\uparrow}]\hat{n}_{\downarrow}
+\hat{n}_{\uparrow}[\Delta_{\downarrow}^{*}
\hat{c}^{\dagger}_{\downarrow}
+\Delta_{\downarrow}\hat{c}_{\downarrow}]),\nonumber\\
\hat{\omega}_{1} &\stackrel{\rm{def}}{=} &
\frac{|\Delta_{\uparrow}|^2}{2}
\left[\hat{c}_{\uparrow},\hat{c}^{\dagger}_{\uparrow}\right]
+\frac{|\Delta_{\downarrow}|^2}{2}
\left[\hat{c}_{\downarrow},\hat{c}^{\dagger}_{\downarrow}\right]
+\Delta_{\downarrow}\Delta_{\uparrow}^{*}\hat{c}_{\downarrow}
\hat{c}^{\dagger}_{\uparrow}
+\Delta_{\uparrow}\Delta_{\downarrow}^{*}\hat{c}_{\uparrow}
\hat{c}^{\dagger}_{\downarrow}
\label{R6}
\end{eqnarray} 

	The commutator of $\hat{l}_{-}$ with the Hamiltonian
yields the final relevant operator of the present set,
namely $\hat{\omega}_{2}$, 
\begin{equation}
\left[\hat{H},\hat{l}_{-}\right] = 
i(\alpha + U)\hat{l}_{+}-i\tau\hat{\omega}_{2},
\label{R7}  
\end{equation}\\[0.3ex]
$\hat{\omega}_{2}$ is given by the following expression
\begin{eqnarray}
\hat{\omega}_{2} &\stackrel{\rm{def}}{=} &
2(\frac{|\Delta_{\uparrow}|^2}{2}
\left[\hat{c}_{\uparrow},\hat{c}^{\dagger}_{\uparrow}\right]
\hat{n}_{\downarrow}
+\frac{|\Delta_{\downarrow}|^2}{2}\hat{n}_{\uparrow}
\left[\hat{c}_{\downarrow},\hat{c}^{\dagger}_{\downarrow}\right]
+\Delta_{\downarrow}\Delta_{\uparrow}^{*}\hat{c}_{\downarrow}
\hat{c}^{\dagger}_{\uparrow}
+\Delta_{\uparrow}\Delta_{\downarrow}^{*}\hat{c}_{\uparrow}
\hat{c}^{\dagger}_{\downarrow}).
\label{R8}
\end{eqnarray} 
The three remaining commutation relations required
to close the algebra can be expressed entirely in terms of 
operators already defined. These read
\begin{eqnarray}
\left[\hat{H},\hat{l}_{+}\right] &=& 
-i(\alpha + U)\hat{l}_{-},\nonumber\\
\left[\hat{H},\hat{\omega}_{1}\right] &=& 
i 2 |\Delta|^2 \tau \hat{p},\nonumber\\
\left[\hat{H},\hat{\omega}_{2}\right] &=& 
i 8 |\Delta|^2 \tau \hat{l}_{-}.
\label{R9}
\end{eqnarray}

	Thus we have a set of seven relevant operators,
namely $\hat{n}$,~ $\hat{x}$,~$\hat{p}$,
$\hat{l}_{-}$,~$\hat{l}_{+}$,~$\hat{\omega}_{1}$
and $\hat{\omega}_{2}$ which close the algebra
as is clear from Eqs.~\ref{R1},~\ref{R3},~\ref{R5},
~\ref{R7} and \ref{R9}. Using Eqs.~\ref{F5} and
\ref{F6}, the
evolution equations for the present set of
relevant operators immediately follow and
are 
\begin{eqnarray}
\frac{d<\!\hat{n}\!>_{t}}{dt} & = &
\frac{\tau}{\hbar}<\!\hat{p}\!>_{t}
,\label{R10}\\[1.0ex]
\frac{d<\!\hat{x}\!>_{t}}{dt} & = &
\frac{\alpha}{\hbar}<\!\hat{p}\!>_{t}
+\frac{U}{\hbar}<\!\hat{l}_{-}\!>_{t}
,\label{R11}\\[1.0ex]
\frac{d<\!\hat{p}\!>_{t}}{dt} & = &
-\frac{\alpha}{\hbar}<\!\hat{x}\!>_{t}
-\frac{U}{\hbar}<\!\hat{l}_{+}\!>_{t}
+\frac{4\tau}{\hbar}<\!\hat{\omega}_{1}\!>_{t}
,\label{R12}\\[1.0ex]
\frac{d<\!\hat{l}_{+}\!>_{t}}{dt} & = &
\frac{(\alpha+U)}{\hbar}<\!\hat{l}_{-}\!>_{t}
,\label{R13}\\[1.0ex]
\frac{d<\!\hat{l}_{-}\!>_{t}}{dt} & = &
-\frac{(\alpha+U)}{\hbar}<\!\hat{l}_{-}\!>_{t}
+\frac{\tau}{\hbar}<\!\hat{\omega}_{2}\!>_{t}
,\label{R14}\\[1.0ex]
\frac{d<\!\hat{\omega}_{1}\!>_{t}}{dt} & = &
-\frac{2|\Delta|^2 \tau}{\hbar}<\!\hat{p}\!>_{t}
,\label{R15}\\[1.0ex]
\frac{d<\!\hat{\omega}_{2}\!>_{t}}{dt} & = &
-\frac{8|\Delta|^2 \tau}{\hbar}<\!\hat{l}_{-}\!>_{t}
,\label{R16}\\[1.0ex]
\end{eqnarray}\\[0.3ex]

	The magnetic field term modifies 
the HH by a simple looking term, viz $\hat{N}$,
as is immediately apparent from the Hamiltonian form in
Eq.~\ref{H14}. We observe that $\hat{N}$ differs by
a negative sign between the number operators of spin-up
and spin-down states from $\hat{n}$. This observation
leads us to expect that like $\hat{n}$, $\hat{N}$
when commuted with the Hamiltonian will lead to
a set of relevant operators parallel to the ones
obtained in case of $\hat{n}$.
 
\begin{equation}
\left[\hat{H},\hat{N}\right] = -i \tau \hat{P} ,
\label{R17}  
\end{equation}\\[0.3ex]
where $\hat{P}$ is given by
\begin{eqnarray}
\hat{P} &\stackrel{\rm{def}}{=} & 
i(\Delta_{\uparrow}^{*}\hat{c}^{\dagger}_{\uparrow}
-\Delta_{\downarrow}^{*}\hat{c}^{\dagger}_{\downarrow}
-\Delta_{\uparrow}\hat{c}_{\uparrow}
+\Delta_{\downarrow}\hat{c}_{\downarrow}).
\label{R18}
\end{eqnarray} 

\begin{equation}
\left[\hat{H},\hat{X}\right] = -i \alpha \hat{P}-iU\hat{L}_{-} 
-ih\hat{p}+ i 4 \tau \hat{\Omega}_{1}
\label{R19}  
\end{equation}\\[0.3ex]
where $\hat{X}$,~$\hat{L}_{-}$,~$\hat{\Omega}_{1}$ are
given by 
\begin{eqnarray}
\hat{X} &\stackrel{\rm{def}}{=} & 
\Delta_{\uparrow}^{*}\hat{c}^{\dagger}_{\uparrow}
-\Delta_{\downarrow}^{*}\hat{c}^{\dagger}_{\downarrow}
+\Delta_{\uparrow}\hat{c}_{\uparrow}
-\Delta_{\downarrow}\hat{c}_{\downarrow},\nonumber\\
\hat{L}_{-} &\stackrel{\rm{def}}{=} & 
i([\Delta_{\uparrow}^{*}\hat{c}^{\dagger}_{\uparrow}
-\Delta_{\uparrow}\hat{c}_{\uparrow}]\hat{n}_{\downarrow}
-\hat{n}_{\uparrow}[\Delta_{\downarrow}^{*}
\hat{c}^{\dagger}_{\downarrow}
-\Delta_{\downarrow}\hat{c}_{\downarrow}]),\nonumber\\
\hat{\Omega}_{1} &\stackrel{\rm{def}}{=} &
i(\Delta_{\uparrow}^{*}\Delta_{\downarrow}^{*}
\hat{c}_{\uparrow}^{\dagger}\hat{c}^{\dagger}_{\downarrow}
+\Delta_{\uparrow}\Delta_{\downarrow}
\hat{c}_{\uparrow}\hat{c}_{\downarrow}
+\Delta_{\uparrow}\Delta_{\downarrow}^{*}
\hat{c}_{\uparrow}\hat{c}^{\dagger}_{\downarrow}
+\Delta_{\uparrow}^{*}\Delta_{\downarrow}
\hat{c}^{\dagger}_{\uparrow}\hat{c}_{\downarrow}).
\label{R20}
\end{eqnarray} 
\begin{equation}
\left[\hat{H},\hat{P}\right] = i \alpha \hat{X}+iU\hat{L}_{+} 
+ih\hat{x}-i 4 \tau \hat{\Omega}_{2},
\label{R21}  
\end{equation}\\[0.3ex]
where ~$\hat{L}_{+}$,~$\hat{\Omega}_{2}$ read
\begin{eqnarray}
\hat{L}_{+} &\stackrel{\rm{def}}{=} & 
([\Delta_{\uparrow}^{*}\hat{c}^{\dagger}_{\uparrow}
+\Delta_{\uparrow}\hat{c}_{\uparrow}]\hat{n}_{\downarrow}
-\hat{n}_{\uparrow}[\Delta_{\downarrow}^{*}
\hat{c}^{\dagger}_{\downarrow}
+\Delta_{\downarrow}\hat{c}_{\downarrow}]),\nonumber\\
\hat{\Omega}_{2} &\stackrel{\rm{def}}{=} &
(-\frac{|\Delta_{\uparrow}|^2}{2}
\left[\hat{c}_{\uparrow},\hat{c}^{\dagger}_{\uparrow}\right]
+\frac{|\Delta_{\downarrow}|^2}{2}
\left[\hat{c}_{\downarrow},\hat{c}^{\dagger}_{\downarrow}\right]
+\Delta_{\uparrow}^{*}\Delta_{\downarrow}^{*}
\hat{c}_{\uparrow}^{\dagger}\hat{c}^{\dagger}_{\downarrow}
+\Delta_{\downarrow}\Delta_{\uparrow}
\hat{c}_{\downarrow}\hat{c}_{\uparrow}).
\label{R22}
\end{eqnarray} 
 
The commutators of ~$\hat{L}_{+}$ and ~$\hat{L}_{-}$
with the Hamiltonian are 
\begin{equation}
\left[\hat{H},\hat{L}_{+}\right] = -i (\alpha +U)\hat{L}_{-} 
-ih\hat{l}_{-}- i 2 \tau \hat{\Omega}_{1},
\label{R23}  
\end{equation}
and
\begin{equation}
\left[\hat{H},\hat{L}_{-}\right] = i (\alpha +U)\hat{L}_{+} 
+ih\hat{l}_{+}+i 2 \tau \hat{\Omega}_{2}.
\label{R24}  
\end{equation}

The commutators of $\hat{n}$,~$\hat{x}$,~$\hat{p}$,
with the Hamiltonian in presence of magnetic field
are

\begin{eqnarray}
\left[\hat{H},\hat{n}\right] &=& -i \tau \hat{p} ,\nonumber\\
\left[\hat{H},\hat{x}\right] &=& -i \alpha \hat{p}
-iU\hat{l}_{-}-ih\hat{P} ,\nonumber\\
\left[\hat{H},\hat{p}\right] &=& i \alpha \hat{x}+iU\hat{l}_{+} 
-i 4 \tau \hat{\omega}_{1}+i h \hat{X},\nonumber\\
\left[\hat{H},\hat{l}_{+}\right] &=& 
-i(\alpha + U)\hat{l}_{-}-ih \hat{L}_{-},\nonumber\\
\left[\hat{H},\hat{l}_{-}\right] &=& 
i(\alpha + U)\hat{l}_{+}-i\tau\hat{\omega}_{2}
+ih\hat{L}_{+},\nonumber\\
\left[\hat{H},\hat{\omega}_{1}\right] &=& 
i 2 |\Delta|^2 \tau \hat{p}-i2 h \hat{\Omega}_{3},\nonumber\\
\left[\hat{H},\hat{\omega}_{2}\right] &=& 
i 8 |\Delta|^2 \tau \hat{l}_{-}-i 4 h \hat{\Omega}_{3}.
\label{R25}  
\end{eqnarray}\\[0.3ex]
If we set $h$ to zero we recover the equations
obtained before, which provides a check on our 
calculations.
$\Omega_{3}$ and $\Omega_{4}$ are defined as
\begin{eqnarray}
\hat{\Omega}_{3} &\stackrel{\rm{def}}{=} &
i(\Delta_{\downarrow}^{*}\Delta_{\uparrow}
\hat{c}_{\downarrow}^{\dagger}\hat{c}_{\uparrow}
+\Delta_{\downarrow}\Delta_{\uparrow}^{*}
\hat{c}_{\downarrow}\hat{c}^{\dagger}_{\uparrow}),\nonumber\\
\hat{\Omega}_{4} &\stackrel{\rm{def}}{=} &
\Delta_{\downarrow}^{*}\Delta_{\uparrow}
\hat{c}_{\downarrow}^{\dagger}\hat{c}_{\uparrow}
-\Delta_{\downarrow}\Delta_{\uparrow}^{*}
\hat{c}_{\downarrow}\hat{c}^{\dagger}_{\uparrow}.
\label{R26}
\end{eqnarray} 
In addition we define two more operators
$\Omega_{5}$ and $\Omega_{6}$
\begin{eqnarray}
\hat{\Omega}_{5} &\stackrel{\rm{def}}{=} &
\Delta_{\uparrow}^{*}\Delta_{\downarrow}^{*}
\hat{c}_{\uparrow}^{\dagger}\hat{c}_{\downarrow}^{\dagger}
+\Delta_{\downarrow}\Delta_{\uparrow}
\hat{c}_{\downarrow}\hat{c}_{\uparrow},\nonumber\\
\hat{\Omega}_{6} &\stackrel{\rm{def}}{=} &
i(\Delta_{\uparrow}^{*}\Delta_{\downarrow}^{*}
\hat{c}_{\uparrow}^{\dagger}\hat{c}_{\downarrow}^{\dagger}
-\Delta_{\downarrow}\Delta_{\uparrow}
\hat{c}_{\downarrow}\hat{c}_{\uparrow}).
\label{R27}
\end{eqnarray} 
It is clear from the definitions of $\hat{\Omega}_{1}$,
$\hat{\Omega}_{3}$ and $\hat{\Omega}_{6}$ given respectively
in ~\ref{R20},~\ref{R26} and \ref{R27} that
\begin{equation}
\hat{\Omega}_{6}=\hat{\Omega}_{1}+\hat{\Omega}_{3}.
\label{R28}
\end{equation}
Eq.~\ref{R28} provides a check on our calculation
since it implies that once we have independently
calculated the time-evolution equations for
$\hat{\Omega}_{6}$,~$\hat{\Omega}_{1}$,
and ~$\hat{\Omega}_{3}$ they must obey the relation
\begin{equation}
\frac{d<\!\hat{\Omega}_6\!>_{t}}{dt}
=\frac{d<\!\hat{\Omega}_1\!>_{t}}{dt}
+\frac{d<\!\hat{\Omega}_3\!>_{t}}{dt}.
\label{R29}
\end{equation}
Similarly it follows from definitions of $\hat{\Omega}_{2}$
[see ~\ref{R22}], $\hat{\Omega}_{5}$ [see ~\ref{R27}] and the
definitions of $\hat{n}$ and $\hat{N}$ that
\begin{equation}
\hat{\Omega}_{5}=\hat{\Omega}_{2}
+|\tilde{\Delta}|^2\hat{I}-|\tilde{\Delta}|^2\hat{n}
-|\Delta|^2\hat{N}
\label{R30}
\end{equation}
which implies that 
\begin{equation}
\frac{d<\!\hat{\Omega}_5\!>_{t}}{dt}
=\frac{d<\!\hat{\Omega}_2\!>_{t}}{dt}
-|\tilde{\Delta}|^2\frac{d<\!\hat{n}\!>_{t}}{dt}
-|\tilde{\Delta}|^2\frac{d<\!\hat{N}\!>_{t}}{dt}.
\label{R31}
\end{equation}

	The commutators of $\hat{\Omega}_{1}$ through
$\hat{\Omega}_{6}$ with the Hamiltonian are obtained
after some calculation and may be displayed as
\begin{eqnarray}
\left[\hat{H},\hat{\Omega}_1\right] &=& 
i (2\alpha+U)\Omega_{5}
-i~2~\tau [|\tilde{\Delta}|^2\hat{x}-|\Delta|^2 \hat{X}]
+ i~2~h~\Omega_{4},\nonumber\\
\left[\hat{H},\hat{\Omega}_2\right] &=& -i (2\alpha+U)\Omega_{6} 
-2i\tau |\Delta|^2\hat{P} ,\nonumber\\
\left[\hat{H},\hat{\Omega}_3\right] &=& 
i\tau |\tilde{\Delta}|^2 \hat{x}-i\tau |\Delta|^2 \hat{X}
- i 2 h \Omega_{4},\nonumber\\
\left[\hat{H},\hat{\Omega}_{4}\right] &=& 
i \tau |\Delta|^2 \hat{p}- i \tau |\tilde{\Delta}|^2 \hat{P}
-i 2 h \Omega_{3},\nonumber\\
\left[\hat{H},\hat{\Omega}_{5}\right] &=& 
-i (2 \alpha+U) \Omega_{6}
-i\tau |\Delta|^2 \hat{P}
-i\tau |\tilde{\Delta}|^2 \hat{p},\nonumber\\
\left[\hat{H},\hat{\Omega}_{6}\right] &=& 
i (2 \alpha+U) \Omega_{5}
-i \tau |\tilde{\Delta}|^2 \hat{x}
+ i \tau |\Delta|^2 \hat{X}.
\label{R32}  
\end{eqnarray}\\[0.3ex]

	It follows from the above discussion that
we have a set of eighteen relevant operators 
in the presence of the external magnetic field,
namely $\hat{n}$,~ $\hat{x}$,~$\hat{p}$,
$\hat{l}_{-}$,~$\hat{l}_{+}$,~$\hat{\omega}_{1}$
,~$\hat{\omega}_{2}$, $\hat{N}$,~ $\hat{X}$,~$\hat{P}$,
$\hat{L}_{-}$,~$\hat{L}_{+}$,~$\hat{\Omega}_{1}$
,~$\hat{\Omega}_{2}$,~$\hat{\Omega}_{3}$,
~$\hat{\Omega}_{4}$,~$\hat{\Omega}_{5}$ ,
and ~$\hat{\Omega}_{6}$ which close the algebra
as is clear from Eqs.~\ref{R1},~\ref{R3},~\ref{R5},
~\ref{R7},~\ref{R9},~\ref{R17},~\ref{R19},~\ref{R21},
~\ref{R24},~\ref{R23}  and \ref{R32}. However not all 
of the operators are independent as is clear from 
the relations given in Eqs.~\ref{R28} and ~\ref{R30}.
Using Eqs.~\ref{F5} and
\ref{F6}, the evolution equations for the present 
set of relevant operators can be written as
\begin{eqnarray}
\frac{d<\!\hat{n}\!>_{t}}{dt} & = &
\frac{\tau}{\hbar}<\!\hat{p}\!>_{t}
,\label{R40}\\[1.0ex]
\frac{d<\!\hat{x}\!>_{t}}{dt} & = &
\frac{\alpha}{\hbar}<\!\hat{p}\!>_{t}
+\frac{U}{\hbar}<\!\hat{l}_{-}\!>_{t}
+\frac{h}{\hbar}<\!\hat{P}\!>_t
,\label{R41}\\[1.0ex]
\frac{d<\!\hat{p}\!>_{t}}{dt} & = &
-\frac{\alpha}{\hbar}<\!\hat{x}\!>_{t}
-\frac{U}{\hbar}<\!\hat{l}_{+}\!>_{t}
+\frac{4\tau}{\hbar}<\!\hat{\omega}_{1}\!>_{t}
-\frac{h}{\hbar} <\!\hat{X}\!>_{t}
,\label{R42}\\[1.0ex]
\frac{d<\!\hat{l}_{+}\!>_{t}}{dt} & = &
\frac{(\alpha+U)}{\hbar}<\!\hat{l}_{-}\!>_{t}
+\frac{h}{\hbar} <\!\hat{L}_{-}\!>_{t}
,\label{R43}\\[1.0ex]
\frac{d<\!\hat{l}_{-}\!>_{t}}{dt} & = &
-\frac{(\alpha+U)}{\hbar}<\!\hat{l}_{-}\!>_{t}
+\frac{\tau}{\hbar}<\!\hat{\omega}_{2}\!>_{t}
-\frac{h}{\hbar} <\!\hat{L}_{+}\!>_{t}
,\label{R44}\\[1.0ex]
\frac{d<\!\hat{\omega}_{1}\!>_{t}}{dt} & = &
-\frac{2|\Delta|^2 \tau}{\hbar}<\!\hat{p}\!>_{t}
+2\frac{h}{\hbar} <\!\hat{\Omega}_{3}\!>_{t}
,\label{R45}\\[1.0ex]
\frac{d<\!\hat{\omega}_{2}\!>_{t}}{dt} & = &
-\frac{8|\Delta|^2 \tau}{\hbar}<\!\hat{l}_{-}\!>_{t}
+4\frac{h}{\hbar} <\!\hat{\Omega}_{3}\!>_{t}
,\label{R46}\\[1.0ex]
\frac{d<\!\hat{N}\!>_{t}}{dt} & = &
\frac{\tau}{\hbar}<\!\hat{P}\!>_{t}
,\label{R47}\\[1.0ex]
\frac{d<\!\hat{X}\!>_{t}}{dt} & = &
\frac{\alpha}{\hbar}<\!\hat{P}\!>_{t}
+\frac{U}{\hbar}<\!\hat{L}_{-}\!>_{t}
+\frac{h}{\hbar}<\!\hat{p}\!>_t
-4 \frac{\tau}{\hbar}<\!\hat{\Omega}_{1}\!>_t
,\label{R48}\\[1.0ex]
\frac{d<\!\hat{P}\!>_{t}}{dt} & = &
-\frac{\alpha}{\hbar}<\!\hat{X}\!>_{t}
-\frac{U}{\hbar}<\!\hat{L}_{+}\!>_{t}
-\frac{h}{\hbar} <\!\hat{x}\!>_{t}
+4 \frac{\tau}{\hbar}<\!\hat{\Omega}_{2}\!>_{t}
,\label{R49}\\[1.0ex]
\frac{d<\!\hat{L}_{+}\!>_{t}}{dt} & = &
\frac{(\alpha+U)}{\hbar}<\!\hat{L}_{-}\!>_{t}
+\frac{h}{\hbar} <\!\hat{l}_{-}\!>_{t}
+2\frac{\tau}{\hbar}<\!\hat{\Omega}_{1}\!>_{t}
,\label{R50}\\[1.0ex]
\frac{d<\!\hat{L}_{-}\!>_{t}}{dt} & = &
-\frac{(\alpha+U)}{\hbar}<\!\hat{L}_{+}\!>_{t}
-\frac{h}{\hbar} <\!\hat{l}_{+}\!>_{t}
-2\frac{\tau}{\hbar}<\!\hat{\Omega}_{2}\!>_{t}
,\label{R51}\\[1.0ex]
\frac{d<\!\hat{\Omega}_1\!>_{t}}{dt} & = &
-\frac{(2 \alpha+U)}{\hbar}<\!\hat{\Omega}_{5}\!>_{t}
-2 \frac{h}{\hbar} <\!\hat{\Omega}_{4}\!>_{t}
+2\frac{\tau}{\hbar} [|\tilde{\Delta}|^2 <\!\hat{x}\!>_{t}
-|\Delta|^2 <\!\hat{X}\!>_{t}]
,\label{R52}\\[1.0ex]
\frac{d<\!\hat{\Omega}_2\!>_{t}}{dt} & = &
+\frac{(2\alpha+U)}{\hbar}<\!\hat{\Omega}_{6}\!>_{t}
+2\frac{\tau}{\hbar}|\Delta|^2 <\!\hat{P}\!>_{t}
,\label{R53}\\[1.0ex]
\frac{d<\!\hat{\Omega}_3\!>_{t}}{dt} & = &
-\frac{\tau}{\hbar} |\tilde{\Delta}|^2  <\!\hat{x}\!>_{t}
+\frac{\tau}{\hbar} |\Delta|^2  <\!\hat{X}\!>_{t}
+2 \frac{h}{\hbar}  <\!\Omega_{4}\!>_{t}
,\label{R54}\\[1.0ex]
\frac{d<\!\hat{\Omega}_4\!>_{t}}{dt} & = &
-\frac{\tau}{\hbar} |\Delta|^2  <\!\hat{p}\!>_{t}
+\frac{\tau}{\hbar} |\tilde{\Delta}|^2  <\!\hat{P}\!>_{t}
+2 \frac{h}{\hbar}  <\!\Omega_{3}\!>_{t}
,\label{R55}\\[1.0ex]
\frac{d<\!\hat{\Omega}_5\!>_{t}}{dt} & = &
+\frac{(2 \alpha+U)}{\hbar} <\!\Omega_{6}\!>_{t}
-\frac{\tau}{\hbar} |\tilde{\Delta}|^2  <\!\hat{p}\!>_{t}
+\frac{\tau}{\hbar} |\Delta|^2 <\!\hat{P}\!>_{t}
,\label{R56}\\[1.0ex]
\frac{d<\!\hat{\Omega}_6\!>_{t}}{dt} & = &
-\frac{(2 \alpha+U)}{\hbar} <\!\Omega_{5}\!>_{t}
+\frac{\tau}{\hbar} |\tilde{\Delta}|^2  <\!\hat{x}\!>_{t}
- \frac{\tau}{\hbar} |\Delta|^2 <\!\hat{X}\!>_{t}
,\label{R57}\\[1.0ex]
\end{eqnarray}\\[0.3ex]
To check the identity given in ~\ref{R29} we add
Eqs.~\ref{R52} and ~\ref{R54} and see if the sum
of these two equations agrees with Eq.~\ref{R57}.
We can immediately see that indeed this is the case.
Likewise using Eqs.~\ref{R40},~\ref{R47},~\ref{R53}
and ~\ref{R56} we can see that the relation given
in Eq.~\ref{R31} holds. Thus we have an independent
check of our stated relations. 
 
	The following remark is in order in context
of future outlook. The MEP formalism is limited to the 
mean-field approach. However the group theory based
approach of identifying the set of operators which
close the partial Lie algebra under commutation with
the Hamiltonian, is quite general. It is thus tempting 
to go beyond the mean-field formalism and use the set 
of relevant operators and their evolution equations to 
develop a technique which can take into account the 
quantum fluctuations.

\section{Conclusions}
We have given the set of relevant operators and the 
corresponding temporal evolution equations for the 
Hubbard Hamiltonian in the mean field approximation 
in the context of the maximal entropy formalism. 
The mean field approximation has been applied to
the hopping term in the Hubbard term. As intuitively
expected the inclusion of the external magnetic field
leads to much larger set of the relevant operators
than present in the case where the magnetic field
is absent.
\newpage
\section*{Acknowledgments}
S.~Alam would like to thank A.~N.~Proto for discussion on
the relevant operators in context of the Hubbard Hamiltonian.
The work of S.~Alam was supported by the Japan Society for
the Promotion of Science [JSPS]. 

\end{document}